\documentclass[preprint,number,11pt,times,a4paper]{elsarticle}

\usepackage{amssymb}

\usepackage{graphicx}
\usepackage{dcolumn}
\usepackage{bm}
\usepackage{amsmath}
\usepackage{amssymb}
\usepackage{amsfonts}
\usepackage[utf8]{inputenc}
\usepackage{textcomp}
\usepackage{color}
\usepackage{multirow}

\graphicspath{{./figures/}}

\newcommand{\additions}[1]{#1}

\journal{Journal of Colloid and Interface Science}

\begin{document}

\begin{frontmatter}



\title{The electrorheology of suspensions consisting of Na-Fluorohectorite synthetic clay particles in silicon oil}


\author[GR,NTNU]{Y. Méheust\corref{cor1}}
\author[NTNU,Pohang]{K. P. S. Parmar}
\author[NTNU]{B. Schjelderupsen}
\author[NTNU]{J. O. Fossum}

\cortext[cor1]{Corresponding author: {\tt yves.meheust@univ-rennes1.fr}}

\address[GR]{Geosciences Rennes (UMR CNRS 6118), Université Rennes 1, Rennes, F-35042 France}
\address[NTNU]{Department of Physics, Norwegian University of Science and Technology, NTNU, N-7491 Trondheim, Norway}
\address[Pohang]{Eco-friendly Catalysis and Energy Laboratory(NRL), Department of Chemical Engineering, Pohang University of Science and Technology, Pohang 790-784, South Korea}

\begin{abstract}
Under application of an electric field greater than a triggering electric field $E_c \sim 0.4$ kV/mm,  suspensions obtained by dispersing particles of the synthetic clay fluorohectorite in a silicon oil, aggregate into chain- and/or column-like structures parallel to the applied electric field. This micro-structuring  results in a transition in the suspensions' rheological behavior, from a Newtonian-like behavior to a shear-thinning rheology with a significant yield stress. We study this behavior as a function of particle volume fraction and strength of the applied electric field, $E$. The steady shear flow curves (fixed shear strain) are observed to scale onto a master curve with respect to $E$, in a manner similar to what was recently found for suspensions of synthetic laponite clay \cite{parmarLangmuir2008}. In the case of Na-fluorohectorite, the corresponding dynamic yield stress is demonstrated to scale with respect to $E$ as a power law with an exponent $\alpha \sim 1.93$, while the static yield stress inferred from constant shear stress tests exhibits a similar behavior with $\alpha = 1.58$. The suspensions are also studied in the framework of thixotropic fluids: the bifurcation in the rheology behavior when letting the system flow and evolve under a constant applied shear stress is characterized, and a bifurcation yield stress, estimated as the applied shear stress at which viscosity bifurcation occurs, is measured to scale as $E^\alpha$ with $\alpha \sim 0.5$ to $0.6$. All measured yield stresses increase with the particle fraction $\Phi$ of the suspension. For the static yield stress, a scaling law $\Phi^\beta$, with $\beta=0.54$, is found. The results from the three types of measurements are found to be reasonably consistent with each other. Their similarities with-, and discrepancies to- results previously obtained on laponite-oil suspensions are discussed.
\end{abstract}


\end{frontmatter}


\section{\label{sec:introduction}Introduction}

\begin{figure}
\centering
\includegraphics[width=0.45\textwidth]{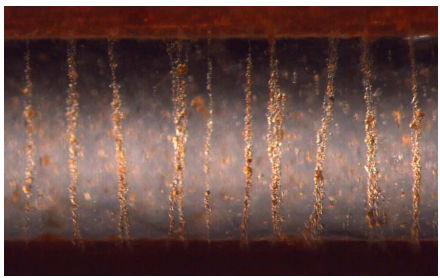}
\caption{\label{fig:ERchains} Microscopy picture of ER chains obtained from applying a $1$kV/mm electric field to a suspension of NaFh clay crystallite aggregates in a silicon oil. The red bands seen in the upper and lower parts of the image are copper electrodes.}
\end{figure}

When suspensions of electrically-polarizable nanolayered clay particles suspended in silicon oil are subjected to an external electric field of the order of $\sim 1$ kV/mm, the particles become polarized, and subsequent dipolar interactions are responsible for aggregating a series of interlinked particles that form chains and columns parallel to the applied field (see Fig. \ref{fig:ERchains}). In the general case \cite{halseyPhysRevLett90,blockLANGMUIR90,halseySCIENCE92,halseySciAmer93,parthasarathyMatSciEng96,weijiaJApplPhys99,HaoAdvMater2001,haoAdvColIntSci2002}, i.e. whatever the nature of the suspensed particles, this kind of structuring stabilizes within some tens of seconds, and disappears when the field is removed. The structuring coincides with a drastic change in the rheological properties of the suspensions, which is why they are called Electro-Rheological (ER) fluids \cite{halseySCIENCE92}; the mechanical behavior of such a system is thus controllable by the applied electric field \cite{halseyPhysRevLett90,blockLANGMUIR90,halseySCIENCE92,halseySciAmer93,parthasarathyMatSciEng96,weijiaJApplPhys99,HaoAdvMater2001,haoAdvColIntSci2002}. Various physical and chemical properties of particles and the suspending liquid may control the behavior of an ER-fluid: Dielectric constant and/or- conductivity of the suspended particles \cite{blockLANGMUIR90,parthasarathyMatSciEng96}, volume fraction of particles \cite{tianMaterLett2003}, frequency and magnitude of the applied electric field  \cite{blockLANGMUIR90,parthasarathyMatSciEng96}. Other factors such as particle geometry \cite{KanuJRheol98,DuanJIntellMatSystemsStructures2000,tianMaterLett2003}, size \cite{shihIntJModPhys1994,TanPhysRevE1999} and polydispersity  \cite{shihIntJModPhys1994,TanPhysRevE1999} also influence the ER-shear stress behavior. Small amounts of additives like water \cite{wongCONF89}, adsorbed and/or- absorbed on the particles, may also play an important  role for certain types of ER fluids.

Recently, we have studied the effect of an external DC electric field ($\sim 1$ kV/mm) on the rheological properties of colloidal suspensions consisting of laponite particles in a silicone oil \cite{parmarLangmuir2008}. The laponite particle aggregates were observed to exhibit the polarization and subsequent self-assembly typical of ER fluids. Without an applied electric field, the steady state shear behavior of such laponite suspensions is Newtonian-like. Under application of an electric field larger than the triggering field, the samples' rheological properties changes dramatically: a significant yield stress was measured, and under continuous shear the fluid exhibited a shear-thinning behavior. We studied the rheological properties, in particular the dynamic and static shear stress, $\tau$, as a function of particle volume fraction $\Phi$, for various strengths of the applied electric field $E$. We showed that the flow curves under continuous shearing can be scaled with respect to both $\Phi$ and $E$, onto a master curve in the form:
\begin{equation}
\label{eq:scaling_law}
 \tau (\Phi, E) = \Phi^\beta E^\alpha f\left ( \frac{\dot{\gamma}}{\Phi^{2/3}\, E^2} \right ) \text{~ ,}
\end{equation}
where $\dot{\gamma}$ is the shear rate, $\alpha$ and $\beta$ are scaling exponents, and $f$ represents the master curve. We demonstrated that Eq.~\eqref{eq:scaling_law} is consistent with simple scaling arguments. We showed that in the laponite case, the shape of this master curve approaches a standard power-law model at high Manson numbers. For our laponite suspensions, both dynamic and static yield stress were shown to depend on the particle volume fraction and applied electric field as $\Phi^\beta \, E^\alpha$, with $\alpha \sim 1.85$, and $\beta \sim 1.00$ and $1.70$, for the dynamic and static yield stresses respectively. 

The clay particles investigated in the present study belong to the same family as the laponite particles, namely 2:1 clays, whose basic structural unit is a $1$ nm-thick platelet consisting of two tetrahedral silica sheets sandwiching one octahedral silica sheet. 2:1 clays carry a moderate negative surface charge on their plane surfaces. This charge is usually sufficiently large so that individual platelets be able to stack by sharing cations, and at the same time moderate enough so as to allow further intercalation of water molecules into the resulting “decks of cards”-like smectite particles. Natural 2:1 clay particles dispersed in salt solutions have been studied for decades \cite{veldeBOOK92}, and recently there has been a growing activity in the study of complex physical phenomena in synthetic smectites \cite{fossumPHYSICA99}. Much effort has gone into relating the lamellar microstructure of smectite clay-salt water suspensions to their collective interaction and to resulting macroscopic physical properties, such as phase behavior and rheological properties \cite{GabrielJPhysChem96,mourchidLangmuir98,bonnLangmuir99,DiMasiPRE2001,bonnPRL2002,coussotPRL2002,lemaireEurophysLett2002,fossumENERGY2005,fonsecaJApplCryst2007}. Nematic liquid crystalline-like ordering in smectite systems has been characterized by the observation of birefringent domains with defect textures \cite{GabrielJPhysChem96,lemaireEurophysLett2002,fossumENERGY2005}, by X-Ray scattering \cite{DiMasiPRE2001,fonsecaJApplCryst2007,fonsecaPRE2009,hemmenLangmuir2009} and by NMR spectrometry and imaging \cite{deAzevedoLangmuir2007,hemmenLangmuir2009}.

The synthetic 2:1 clay fluorohectorite (FH) is a hectorite clay in which the hydroxyl groups have been replaced by fluorin atoms. Each individual clay platelet is about $1$ nm thick, and its density is $2.8$ \cite{knudsenJApplCryst2003}. Fluorohectorite particles suspended in saline solutions consist of $\sim 80$\textendash $100$ negatively surface-charged clay platelets \cite{kaviratnaJPhysChemSol96,DiMasiPRE2001,dasilvaPRE2002,daSilvaPRB2003} residing on top of each other with intercalated charge-balancing cations X \cite{kaviratnaJPhysChemSol96,fossumPHYSICA99}, such as Na$^+$, Ni$^{2+}$, or Fe$^{3+}$. They have a large aspect ratio: their thickness is $\sim 0.1$ \textmu{}m, while their lateral size can vary from tens of nm up to a few \textmu{}m. This deck-of-card structure of the X-FH clay particles subsists in a weakly-hydrated state. The layer charge density of the individual platelet is 1.2 e$^-/$unit cell \cite{kaviratnaJPhysChemSol96}, which allows the interlayer cations to be exchanged with other types of cations in solution. A fluorohectorite system in which all intercalated cations have been exchanged so as to be of one type X only is denoted X-FH. In the present study, we have used Na-FH, whose half unit cell has a formula Na$_{0.6}$(Mg$_{2.4}$Li$_{0.6}$)Si$_4$O$_{10}$F$_2$. As for all 2:1 clays, water molecules or other polar molecules can be adsorbed or intercalated between interlayer spaces of the platelets, depending upon the environmental conditions such as temperature and relative humidity \cite{kaviratnaJPhysChemSol96,dasilvaPRE2002,daSilvaPRB2003,knudsenJApplCryst2003,knudsenPHYSICA2004,lovollPhysicaB2005,meheustClayScience2005}.  When observed in a scanning electron microscope (SEM), untreated sodium fluorohectorite (Na-FH) clay powder display particles with edges and surfaces that are very rough and they are highly anisotropic in their shape, and highly polydisperse in size along their lateral directions. The lateral size of such clay particles can be further reduced by fine grinding of its powder.

Recently, we demonstrated that X-FH particles suspended in silicone oil display ER structuring \cite{fossumEPL2006}. We investigated the chain or-column like-structures of X-FH particles \cite{fossumEPL2006} using synchrotron X-ray scattering. We determined the structural morphology (preferred orientations of the X-FH particles) of the chain- or column-like structures and the direction of the induced dipoles in X-FH particles. We pointed to the possible migrations of surface charges and possible movement of interlayer cations X of X-FH as main factors responsible for their polarization in an external DC electric field. 

In the present report we complement the previous structural studies with results from rheometry measurements under steady shear, following the method previously used in our study of electro-rheological laponite suspensions \cite{parmarLangmuir2008}.

\section{Physical background -- The rheology of ER fluids}

\subsection{Rheology of ER suspensions when no external electric field is applied}

Without an electric field, the ideal steady state shear behavior of an ER fluid is Newtonian-like \cite{KlingenbergLangmuir1990} and an increase of its particle volume fraction increases the magnitude of its viscosity $\eta$. The $\Phi$-dependent viscosity of an ER fluid of monodisperse spherical particles suspended in Newtonian-liquid ($\Phi <10$\%) can be approximated by the Batchelor relation and the Krieger-Dougherty relation \cite{macoskoBOOK94}. These relations account for the interactions between the particles themselves and between particles and the surrounding liquid. One empirical equation which has been found to account for the viscosity of monodisperse and polydisperse sphere suspensions in a range of particle fraction up to $50$\%, has been suggested by Chong, Christiansen and Baer \cite{chongJApplPolymSci71}. The Batchelor and Krieger-Dougherty relations can strictly only be used to describe the viscosity of concentrated colloidal suspensions of isotropic and monodisperse non-charged particles, but the Chong-Christiansen-Baer relation \cite{chongJApplPolymSci71} has been found to be an appropriate description for concentrated suspensions of kaolinite clay \cite{coussotPRL1995}, a system that consists of  anisotropic and polydisperse particles.

In our recent study of oil suspensions of synthetic laponite we found that the following generalized Krieger-Dougherty relation can be used to fit the zero field viscosity $\eta$ \cite{parmarLangmuir2008} up to $\Phi = 50$ \%:
\begin{equation}
\label{eq:viscosity_model_by_chong_etal}
\eta = \eta_0 \: \left ( 1 - \frac{\Phi}{\Phi_\text{max}}\right )^{-\eta_\text{I}\ \Phi_\text{max}} \text{~ ,}
\end{equation}
in which the parameter $\eta_\text{I} = \lim_{\Phi \rightarrow 0} \left [ (\eta/\eta_0 -1)/\Phi \right ]$ is the fluid's intrinsic viscosity, and $\eta_0$ is the viscosity of the carrier fluid, while $\Phi_\text{max}$ is the packing value for the volume fraction, which cannot be exceeded.
As $\Phi$ increases, the various interaction forces such as hydrodynamics forces, dispersion forces, electrostatic forces, and steric effects/polymeric forces also increase.
The net result is an increase in the viscosity of the suspension.

\subsection{Rheology when an external electric field is applied}

Under application of an electric field ($\sim 1$ kV/mm) and under steady state shear, a general ER fluid shows a well-defined yield stress beyond which it tends to be shear-thinning (pseudoplastic) i.e. the viscosity decreases with increasing shear rate. The typical steady-shear behavior of an ER fluid under the influence of an external electric field is generally characterized as a Bingham-like solid \cite{KlingenbergLangmuir1990}.
\additions{This is especially true at large enough shear rates, while at lower shear rates significant deviations are observed, which can be described to some extent by more complex functional relations \cite{zhuJNonNewtonianFluidMech2005,ChoPolymer2005}.}
 The value of the dynamic yield stress is strongly influenced by the rheological model and the shear rate range selected, as it is obtained by extrapolating the curves from steady state shearing measurement to zero shear rate. This value can be significantly different from the static yield stress i.e. a yield stress of a disrupt (no shearing) ER fluid \cite{shenoyBOOK99}. There are several models that attempt to explain the shear stress behavior for various ER fluids, mostly based on the theoretical analysis of disrupt configurations. The polarization theory \cite{halseyPhysRevLett90,MartinJChemPhys1996}, based on the mismatch in dielectric constant between the particles and the suspending liquid predicts that the yield stress is proportional to $E^\alpha$, where $E$ is the magnitude of the applied electric field and the exponent $\alpha$ is equal to $2$. In the polarization theory, the response of the particle polarization and the suspending liquid are assumed to be linearly related to the applied electric field. The conduction theories based on the mismatch in conductivity between the particles and the suspending liquid predict the exponent as $1\le \alpha \le 2$  at both low- and high- volume fractions of particles \cite{andersonPROCEEDING92,attenIntJModPhys1994,foulcJElectrostat94,WuJPhysD1996,WuIntJModernPhys1996}. Also multipole effects are often proposed to account for particle interactions at relatively high concentrations of particles \cite{blockLANGMUIR90,parthasarathyMatSciEng96}.

\subsection{Stress-bifurcation in the rheological behavior}
\label{sec:bifurcation_in_viscosity}

In our previous study of the electrorheology of laponite suspensions \cite{parmarLangmuir2008}, we have shown that ER suspensions can be studied in the framework of thixotropic fluids.
Indeed, their rheology is controlled by a competition between a shear-rejuvenation property and an aging mechanism: the shear-rejuvenation results from the destruction of the ER microstructure, which resists flow-induced particle rearrangement; on the contrary, a suspension left at rest builds up a microstructure in time, which can be considered an aging mechanism, although the particle-particle interactions responsible for it are due to an external electric field.

For all thixotropic yield stress fluids, which may be equivalent to saying for all yield stress fluids \cite{mollerSoftMatter2006}, a consequence of this competition mechanism is that different methods of estimating the static yield stress of a fluidlike substance do not necessarily lead to the same result, as it may depend on the shearing history. 
Coussot and coauthors have proposed a simple phenomenological rheology model \cite{coussotPRL2002,coussotPhysRevLett2002b} accounting for the competition between aging and a shear rejuvenation that is function of the state of the microstructure. In addition to illustrating why the yield stress measured by ``classical'' techniques depends on the history of shear, the model also predicts that there can be shear rates that such fluids cannot accomodate, in other words, that flow localization occurs whenever the fluid is imposed \additions{a given shear rate within the ``forbidden'' range}. Under constant applied shear stress, in contrast, a bifurcation in the flow
curves is observed: if the applied stress is smaller
than a critical value, the microstructure buildup is dominant, and
the viscosity of the fluid eventually goes to infinity; if it is larger,
the viscosity goes to a finite value at large times. Shear rate values that are outside of the asymptotic shear rates observed below this bifurcation threshold cannot be accomodated by the fluid without flow localization.
This rheology in the bifurcation is actually suggested by Moller et al. \cite{mollerSoftMatter2006} as a principle to estimate the yield stress of a thixotropic fluid in an univoque manner: initally forcing flow in the flow cell and subsequently letting the system evolve under a constant shear stress allows determination of whether this shear stress value is below or above the bifurcation threshold. Investigating various applied shear stress values then leads to an accurate determination of the fluid's yield stress. This technique was successfully applied to a laponite-based ER fluid \cite{parmarLangmuir2008}.

\section{Experiments}

\subsection{Sample preparation}
\label{sec:sample_preparation}

As in our previous structural study on X-FH in silicone oil \cite{fossumEPL2006}, we have used synthetic fluorohectorite, originally purchased from Corning Inc. (New York) in powder form and subsequently cation-exchanged to sodium (Na$^+$) \cite{fossumEPL2006}. After dialysis, the Na-fluorohectorite clay was first dried at 100\textcelsius{} for $\sim$10 hours and then grinded in powder form with the help of a mortar and spatula. As suspending liquid in the present studies, we used the silicon oil: Dow Corning 200/100 Fluid (dielectric constant of $2.5$, viscosity of $100$ mPa.s and specific density of $0.973$ at $25$\textcelsius{}) with a small conductivity of about $10$ to $12$ S/m \cite{tangJRheology1995}. Four different ER suspensions of dehydrated Na-FH clay particles with different volume fractions ($\Phi$) were prepared following a protocol previously established with oil-laponite suspensions\cite{parmarLangmuir2008}: (1) removal of water traces in the clay powder and oil by 72 hrs-long heating at $130$\textcelsius{}; (2) mixing of the powders and silicone oil inside glass tubes; (3) cooling of the sealed tubes to room temperature; (4) homogeneization of the suspensions by hand and subsequent ultrasonic bath-shaking for $30$ minutes, at $25$\textcelsius{}. The suspensions were then left to rest so as to remove clay particles with a spherical diameter larger than $\sim 10$ \textmu{}m (according to the Stokes law for particles settling). The remaining suspensions were again hand-shaken about five minutes, three times. 
It should be noted that after removal of the large particles, the volume fraction $\Phi^\ast$ of the clay particles is significantly lower than the initial volume fraction, $\Phi$, of the prepared ER suspensions. Consequently, $\Phi^\ast$ is not precisely known for the measurements.

\additions{The clay powder used in the preparation of the suspensions has been observed to be highly polydisperse. Indeed, TEM imaging of randomly-selected particles from the powders, 
provided a statistics for the typical particle dimension, defined as the square root of the measured basal area of $151$ imaged particles. We obtained surprisingly well-defined power law-shaped cumulative distributions. 
We think that the statistics is sufficient to clearly state that the size probability density function in the powder is a monotonically-decreasing function, hence, that the smaller the particles, the more numerous they are. This sort of size distribution is totally different from that of powders from which most ER systems are made, and for which the particle size is selected in a given range and the size probability density function consequently displays a peaked behavior around the average particle size. On the other hand, the distribution of particle thicknesses (perpendicular to their basal sheet), which can be easily inferred from X-ray diffraction measurements \cite{dasilvaPRE2002}, has previously been found to be peaked around $100$ nm.
}

\additions{
Note however, that once suspended in oil, the clay crystallites tend to aggregate with each other, because their surfaces have more affinity with each other than with the oil. Hence, the ``clay particles'' as they appear in the oil are aggregates of the original clay crystallites.  It is difficult to measure the aggregates' dimensions in situ in the oil, as most nano-imaging techniques (electron-microscopy, AFM) cannot easily be used for objects placed in a viscous liquid environment. These aggregates then evolve by deformation and further aggregation after the electric field has been turned on; they then start being visible under the optical microscope, and even by the naked eye.
}

\subsection{Rheology measurement}
\label{sec:rheology_measurements}

The rheology of fluorohectorite clay suspensions was measured under an applied DC electric field, using a Physica MCR 300 Rotational Rheometer equipped with a coaxial cylinder electrorheological Couette cell (Physica ERD CC/27). The outer cylinder diameter is $14.46$ mm, and the inner is $13.33$ mm. The immersion length is 40 mm and the sample volume is $19.35$ ml. Note that the present protocol of sample preparation (see section \ref{sec:sample_preparation}) requires relatively large amounts of clay powder in order to prepare this volume of oil suspension. All rheological measurements were carried out at constant temperature $25$\textcelsius{}. Two grounding brushes connected to the bob's axis induce an artificial $1$ Pa yield stress in all data, but this value is negligible compared to all yield stress values addressed here.

 The steady shear rheological properties were measured after the suspensions had been pre-sheared at $\dot{\gamma} = 200$~s$^{-1}$ for $60$ s in order to ensure the same initial conditions. In order to determine the static yield stress, a control shear stress (CSS) \additions{experiment} was performed: the applied shear stress was increased linearly (by steps of $2$ Pa) on disrupt ER suspensions that had been in applied the DC electric field for 300 s prior to applying the shear stress.

We also studied the suspensions in the framework of thixotropic yield stress fluids, measuring the viscosity of samples that were initially forced to flow and then applied a constant shear stress. We have presented the principle of the method in section \ref{sec:bifurcation_in_viscosity}. Due to the competition between shear-induced rejuvenation and the electrically-induced formation of electro-rheological chains, the observed asymptotic rheology can fall into either one rheological behavior or another, depending on the magnitude of the applied shear stress. The bifurcation shear stress that separates those two rheological behaviors is taken as the yield stress of the suspension, and defined pratically as the “stress below which no permanent flow occurs”.

\section{Results}

\subsection{Rheology of the suspensions in the absence of external electric field}

\begin{figure}
\centering
\includegraphics[width=0.45\textwidth]{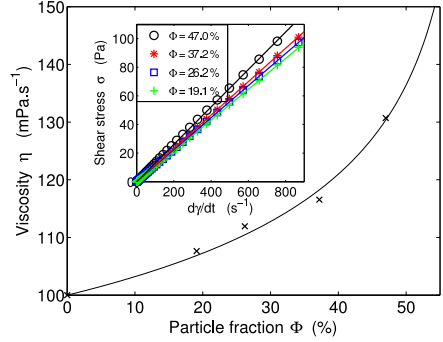}
\caption{\label{fig:figure1} 
Viscosity of the Na-FH suspensions as a function of the volume fraction $\Phi$, at $E \sim 0$. The viscosity values have been obtained from linear fits (shown as plain lines) onto the flow curves (shown as symbols) in the inset.}
\end{figure}
Steady state flow curves (i.e. obtained from Controlled Shear Rate tests) for four different volume fractions ($\Phi$) of Na-FH clay suspensions at zero applied electric field are shown in the inset to Fig.~\ref{fig:figure1}. These suspensions exhibit a Newtonian behavior i.e. a constant viscosity (ratio of the shear stress to the applied shear rate) and no yield stress. The viscosity of the Na-FH clay suspensions increases with the volume fraction of clay particles as illustrated in Fig.~\ref{fig:figure1}. 
Using Eq.~\eqref{eq:viscosity_model_by_chong_etal}, we find  $\eta_I = 0.30$ and $\Phi_\text{m}=61.84$\% as a best fit to the data in Fig.\ref{fig:figure1}. $\Phi_\text{m}$ is expected to be a measure of the maximum packing fraction \cite{chongJApplPolymSci71}. 

As discussed above in section \ref{sec:sample_preparation}, the true $\Phi^\ast < \Phi$ is uncertain in the present case, thus the meaning of $\Phi_\text{m}$ is somewhat lost (the true maximum packing fraction probably being smaller than $\Phi_\text{m}$  in the same manner as $\Phi^\ast$ is overestimated by $\Phi$).

\subsection{CSR tests}

\begin{figure}
\centering
(a)
\includegraphics[width=0.425\textwidth]{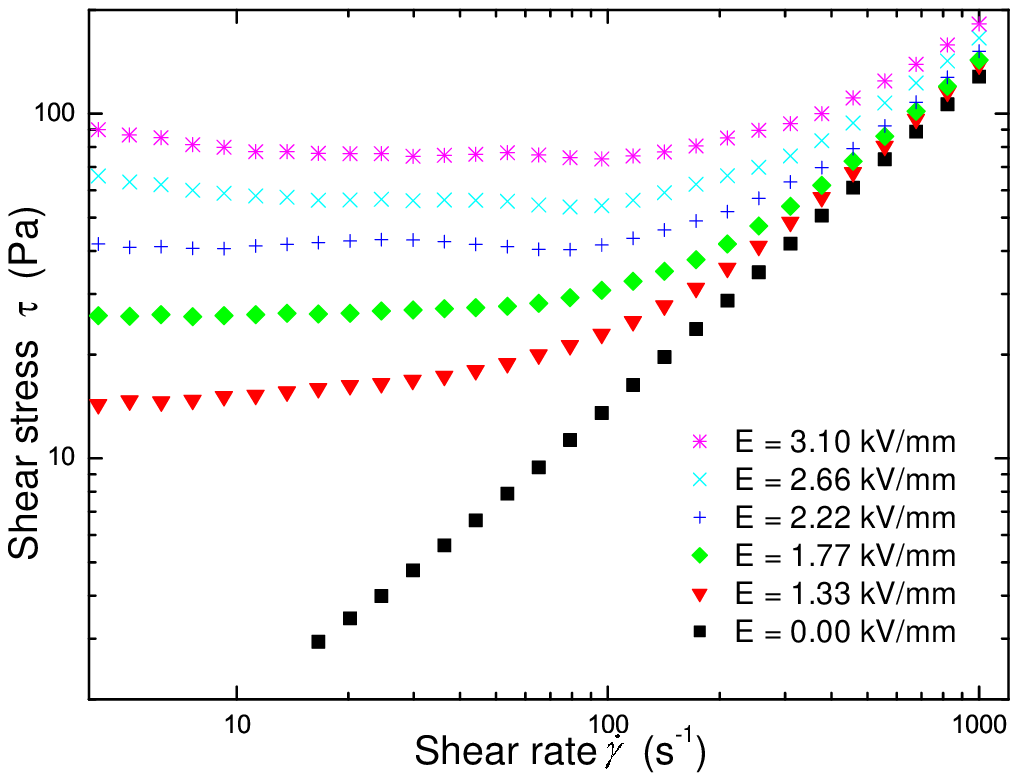}
\vfill
(b)
\includegraphics[width=0.45\textwidth]{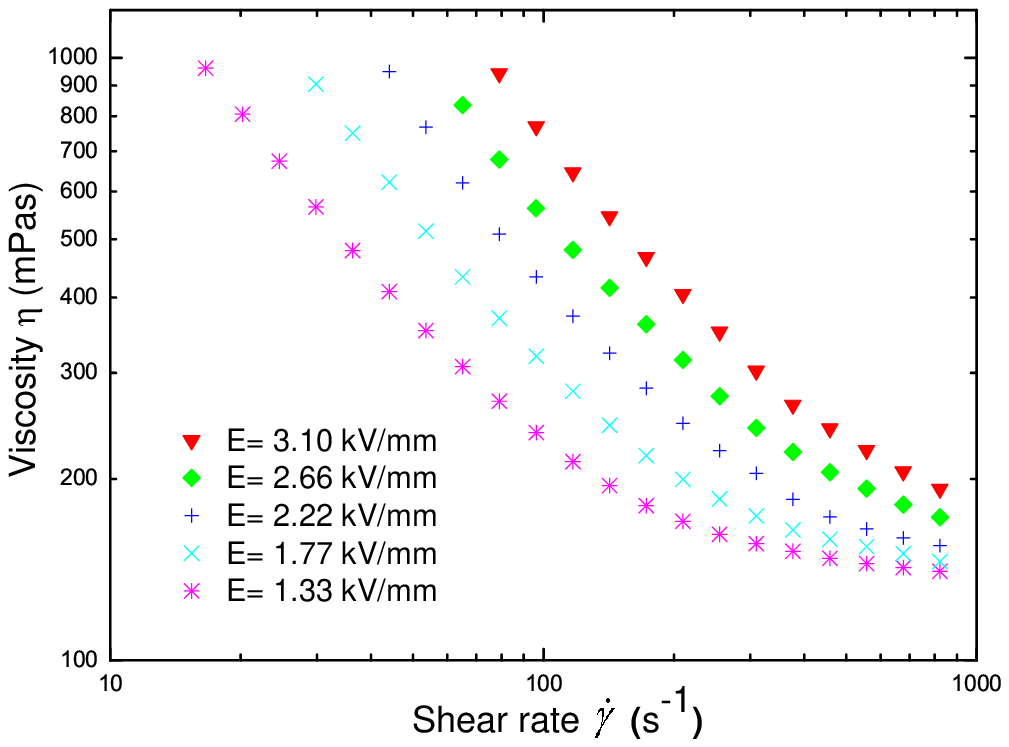}
\\
\caption{\label{fig:figure2}(a) Flow curves of Na-FH suspensions with $\Phi = 37.2$\%, for different strengths of the applied electric fields. (b) Corresponding evolution of the viscosity versus the shear rate.}
\end{figure}
Fig.~\ref{fig:figure2}(a) and \ref{fig:figure2}(b) show the shear stress- and viscosity- curves (obtained from Controlled Shear Rate CSR tests) for a Na-FH clay suspension of volume fraction $\Phi=37.2$~\%, for various magnitudes of the applied DC electric field. Behaviors similar to Fig.~\ref{fig:figure2}(a) and \ref{fig:figure2}(b) are observed for other prepared concentrations of Na-fluorohectorites particles. Increasing the electric field causes an increase in the shear stress (and viscosity) of these suspensions, while the rheology becomes shear thinning over the entire range of shear rates. The flow curves clearly show the yielding behavior of ER suspensions under application of an electric field: the dynamic yield stress $\tau_0^\text{d}$, defined as the limit towards very low shear rates of the measured yield stress, becomes larger as the magnitude of the electric field increases.

\begin{figure}
\centering
\includegraphics[width=0.45\textwidth]{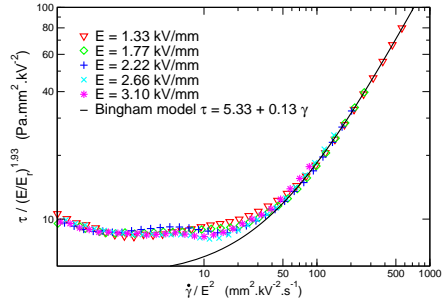}
\caption{\label{fig:figure3} Scaling of the data of Fig.~\ref{fig:figure2}(a) using Eq.~\eqref{eq:scaling_law} with $\alpha = 1.93$, for a particle fraction $\Phi = 37.2$\%. The plain line represents the asymptotic Bingham model. $E_r=1$ kV is the reference electric field value.}
\end{figure}
Following the approach previously used for laponite ER- effects under steady shear \cite{parmarLangmuir2008}, we show in 
Fig.~\ref{fig:figure3} a scaling consistent with Eq.~\eqref{eq:scaling_law} (for a given $\Phi$) applied to the data in Fig.~\ref{fig:figure2}(a). The result is in good agreement with our previous findings for laponite \cite{parmarLangmuir2008}. Fig.~\ref{fig:figure3} corresponds to one concentration $\Phi = 37.2$ \%. The other concentrations that we have investigated display equivalent behaviors with $\alpha \simeq 1.93$. The asymptotic behavior at the highest values of the scaled shear rate, $\dot{\gamma}/E^2$, is Bingham like.

\begin{figure}
\centering
(a)
\includegraphics[width=0.39\textwidth]{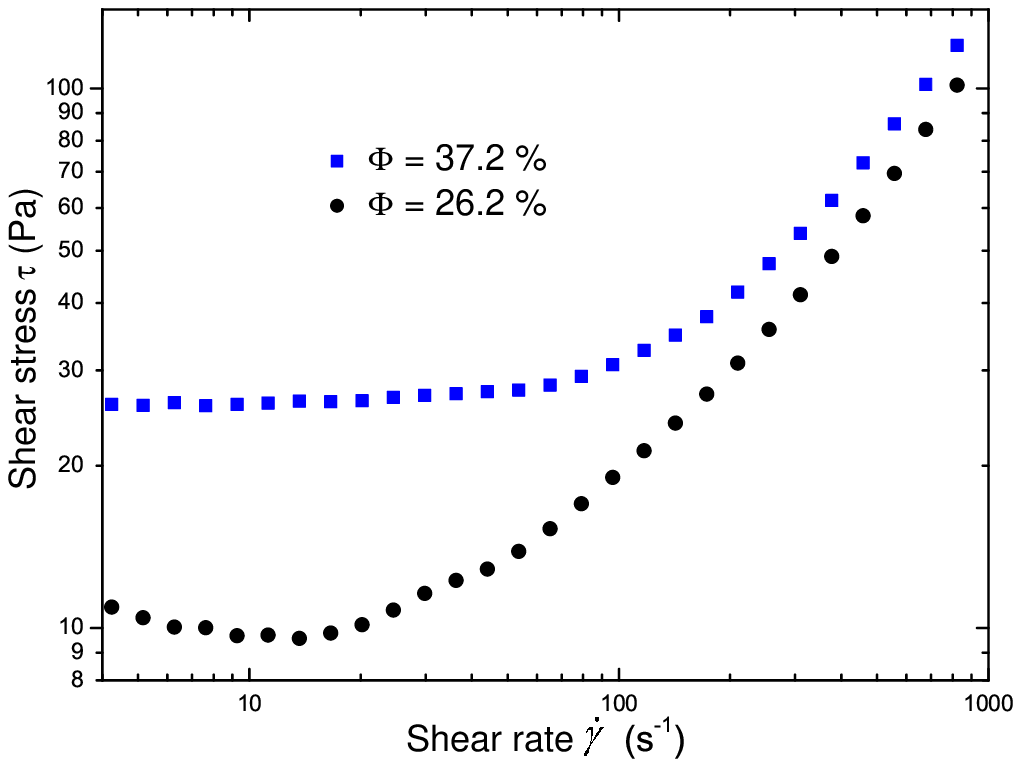}
\vfill
(b)
\includegraphics[width=0.40\textwidth]{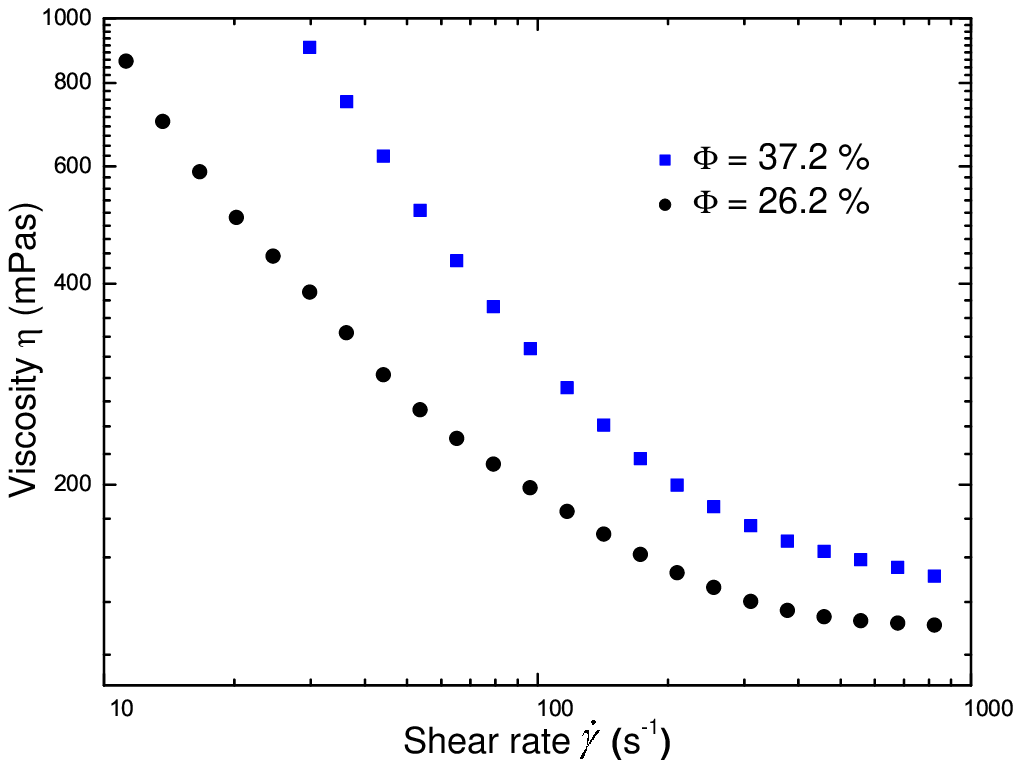}
\caption{\label{fig:figure4}(a) Flow- curves of Na-FH suspension for two different $\Phi$-values at $E\sim 1.77$ kV/mm. (b) Viscosity- curves for the same suspensions.}
\end{figure}
Fig.~\ref{fig:figure4}
shows the effect of the volume fractions ($\Phi$) of clay particles on the flow- and viscosity- curves (Controlled Shear Rate CSR tests) for Na-FH suspensions at a fixed applied electric field $E \sim 1.77$ kV/mm. The shear stress and viscosity increase with the volume fraction ($\Phi$) of clay particles.
In the present case, we have not been able to find a good data collapse and scale the data in Fig.~\ref{fig:figure4} (a) with respect to particle fractions, according to Eq.~\eqref{eq:scaling_law}.

\subsection{CSS tests}

\begin{figure}
\centering
\includegraphics[width=0.45\textwidth]{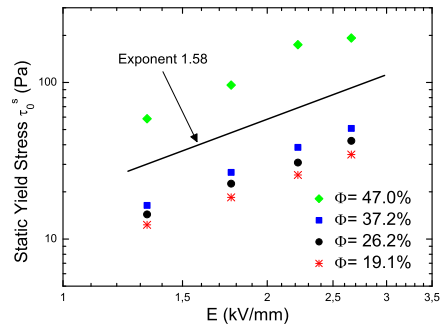}
\caption{\label{fig:figure5} Log-log plot of the static yield stress, $\tau_0^\text{s}$,  plotted versus the strength of the applied DC electric field for different volume fractions of the Na-FH particles.}
\end{figure}

Figure \ref{fig:figure5} shows the static yield stress (obtained from Controlled Shear Stress CSS test) plotted as a function of the applied DC electric field $E$, for different volume fractions ($\Phi$) of the present ER fluids. The static yield stresses, $\tau_0^\text{s}$, increase when the $E$ is increased: all measured static yield stresses are observed to be proportional to $E^\alpha$, with values of exponents $\alpha$ close to $1.58$ for all $\Phi$ values.

\begin{figure} 
\centering
\includegraphics[width=0.45\textwidth]{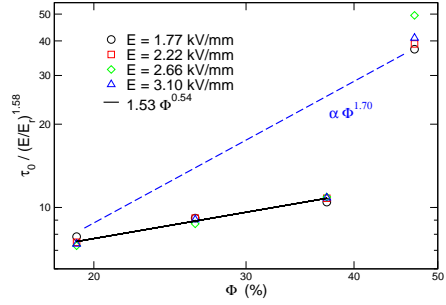}
\caption{\label{fig:figure6} Log-log plot of the scaled static yield  stress versus particle volume fraction
. The dashed line (power law with an exponent $1.70$) represents the behavior of our previously-reported laponite suspensions \cite{parmarLangmuir2008}.}
\end{figure}

In Fig.~\ref{fig:figure6}, we have replotted the yield stress, scaled by $(E/E_r)^{1.58}$, where $E_r=1$ kV is a reference electric field value, as a function of the particle volume fraction. We do obtain a rather satisfying data collapse. For particle fractions up to $37.2$\%, the dependence of the scaled data versus $\Phi$ is a power law with an exponent $0.53$. Between $\Phi=37.2$\% and $\Phi=47.0$, a large “jump” in static yield stress is clearly seen; it is experimentally reproducible for the one sample we have studied at a volume fraction $\Phi = 47.0$\%. At present we cannot explain this “jump”, although we can suggest that it be due to formation of larger effective aggregates above some critical threshold $\Phi$ between $37.2$\% and $47.0$\%.

\subsection{Bifurcation in the rheological behavior}

\begin{figure}
\centering
(a)
\includegraphics[width=0.4\textwidth]{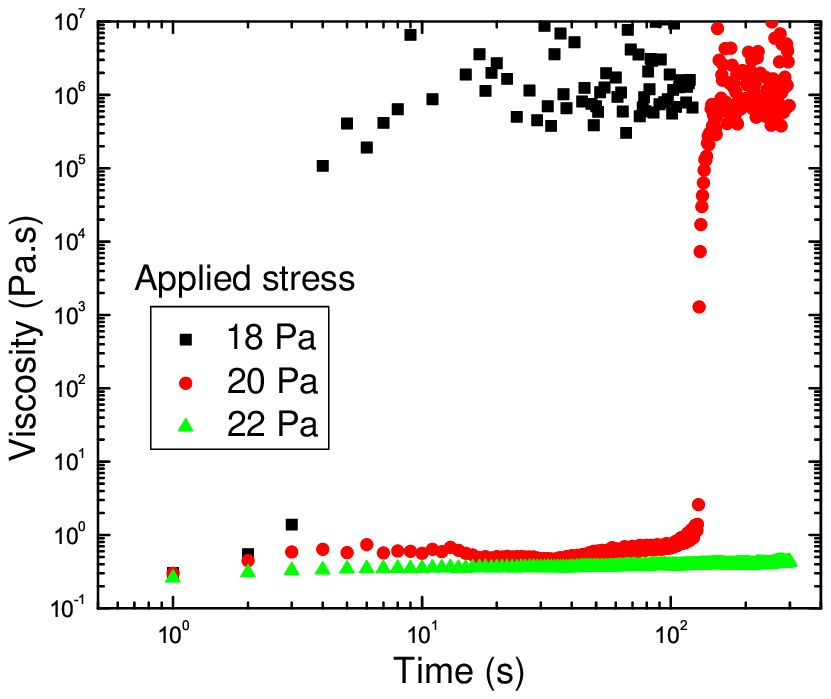}
\vfill
(b)
\includegraphics[width=0.4\textwidth]{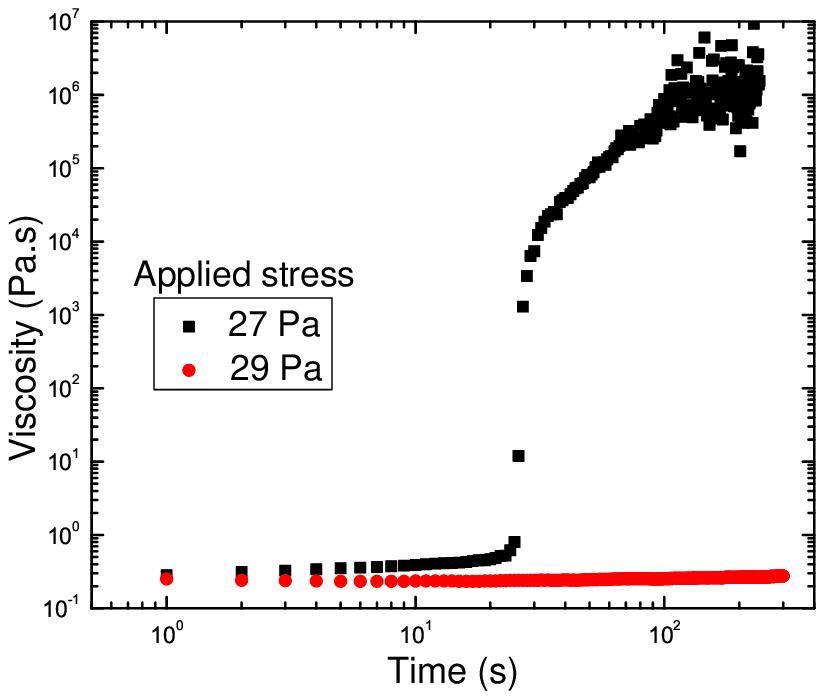}
\\
\caption{\label{fig:bifurcationNaFH} Bifurcation in the rheology of a NaFH suspension of volume fraction $\Phi= 33$\%, under applied electric field strengths of $530$ V.mm$^{-1}$ (a) and $885$ V.mm$^{-1}$ (b).}
\end{figure}

The flow curves obtained when forcing a suspension of volume fraction $\Phi= 33$\% to flow, applying an electric fields of strength $E=530$ V.mm$^{-1}$, and then letting the suspensions evolve under a constant applied stress $\sigma$ are shown in Fig.~\ref{fig:bifurcationNaFH}(a), for various magnitudes of the applied stress. The bifurcation in the rheological behavior occurs for $\tau=\tau_b$ between $20$ and $22$ Pa. The same type of plots are shown for $E=885$ V.mm$^{-1}$ in Fig.~\ref{fig:bifurcationNaFH}(b), leading to $\tau_b$ between $27$ and $29$ Pa. 

\begin{figure}
\centering
\includegraphics[width=0.40\textwidth]{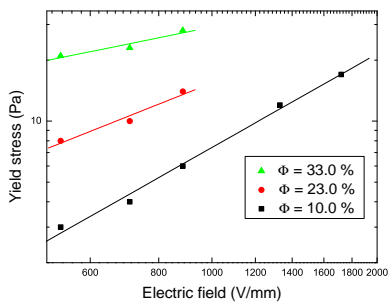}
\caption{\label{fig:bifurcationSigma_vs_E} Bifurcation yield stress $\tau_b$ as a function of the applied electric field, for three different particle volume fractions. The fitted power law exponents are $1.53$, $1.07$ and $0.54$, respectively. This decrease of the exponent with volume fraction is attributed to the growing discrepancy between the measured- and applied- voltage.}
\end{figure}

In contrast to our previous study on laponite particles \cite{parmarLangmuir2008}, the method could not be used for larger strength of the applied electric field because the voltage supply could not maintain the required voltage difference throughout the cell due to leak currents running through the cell. However we were able to infer a bifurcation yield stress for 11 different $(\Phi,E)$ configurations (see table~\ref{tab:results_bifurcation_NaFH}). These measured bifurcation stresses are plotted as a function of the electric field in Fig.~\ref{fig:bifurcationSigma_vs_E}. A power law behavior is observed, but in contrast to what was observed for the CSS and CSR tests, the exponent seem to decrease with increasing volume fraction. We believe that this is due leak currents and to the resulting  discrepancy between the voltage set on the supply and that really applied to the cell. Indeed, this discrepancy is expected to be all the more important as the volume fraction $\Phi$ becomes larger, and therefore a plot of the yield stress versus the nominal electric field will display an exponent all the smaller (with respect to the one that would be observed on a plot versus a measured voltage) as $\Phi$ is larger. Our Physica rheometer is not equipped to measure leak currents that run across the electrorheological cell of the rheometer.
\begin{table}
 \caption{\label{tab:results_bifurcation_NaFH} Critical stresses measured for various electric field strengths and particle fractions. Leak currents prevent accurate measurement for the two larger electric fields and particle fractions.}
\centering
\begin{tabular}{c | c c c}
\hline
\hline
$E$ (V.mm$^{-1}$) & $\Phi = 10$\% & $\Phi = 23$\% & $\Phi = 33$\% \\
\hline
\hline
532 & 3 & 8 & 21 \\
708 & 4 & 10 & 23 \\
885 & 6 & 14 & 28 \\
1330 & 12 & -- & -- \\
1720 & 17 & -- & -- \\
\hline
\hline
\end{tabular}
\end{table}

\subsection{Discussion}

\subsubsection{The electrorheology of Na-FH suspensions}

The three types of measurements performed in this study have shown a power-law dependence versus applied electric field of the yield stress measured on a given suspension.
For reasons summarized in section \ref{sec:bifurcation_in_viscosity} and presented in more details in the introduction to our previous work on electrorheological laponite suspensions \cite{parmarLangmuir2008}, we believe that the best-posed quantity to experimentally characterize the yielding behavior of a thixotropic fluid is the bifurcation yield stress $\tau_b$ such as defined by Møller et al.\cite{mollerSoftMatter2006}. As previously done for laponite suspensions, we have been able to exhibit the rheological bifurcation of the fluorohectorite suspensions. This is a confirmation that such rheological fluids, independently of their composition, can be studied in the framework of thixotropic rheological models. Besides, we could determine a scaling law in the form $\tau_b \propto E^\alpha$ for all studied suspensions, with $\alpha$ exponents that decrease with the particle fraction. In spite of being well-posed theoretically, the bifurcation method is difficult to perform experimentally because it involves lengthy measurements, allowing phenomena such as sedimentation to occur. In this case, we believe that leak currents are the main reasons for the exponent $\beta$ being dependent on $\Phi$ in Fig.~\ref{fig:bifurcationSigma_vs_E}, as the electric field felt by the particles is weakened with respect to the field applied to the ER cell, and this all the more as the average inter-particle distance is small.

This interpretation is consistent with the value $\alpha=1.58$ measured for the static yield stress $\tau_0^\text{s}$ in the range $1.3<E<2.7$ kV (Fig.~\ref{fig:figure5}), which is slightly larger than the $\alpha = 1.53$ measured for $\tau_b$ at $\Phi=10.0$\%. The static yield stress classicly measured with disruption (CSS) tests seems to be consistent with the bifurcation yield stress here. Besides, CSS tests are little affected by sedimentation because they involve a measurement over a short time range, starting with a situation in which all particles are ``frozen'' by the electric field. For particle fractions in the range $19.1\%\leq\Phi\leq37.2\%$, they provide a power-law  dependence of $\tau_0^\text{s} \propto \Phi^\beta$, with $\beta=0.54$ (Fig.~\ref{fig:figure6}).

From the bifurcation tests and CSR tests, we infer a scaling of the suspension yielding behavior in the form
\begin{equation}
 \tau \propto E^\alpha \, \Phi ^\beta \text{~,}
\end{equation}
with $\alpha = 1.55 \pm 0.50$ and $\beta = 0.54\pm 0.05$. When confronted to existing models, the value for $\alpha$ suggests that interaction forces between polarized clay particles are governed by their conductivity mismatch to that of the silicone oil, thus indicating that the polarization of such clay particles in an applied external DC electric field is partly caused by migration of surface charges. In a previous study, we had already shown that the movement of interlayer cations (Na$^+$) sandwiched inside the nano-stacked fluorohectorite particles also plays a role in the polarization \cite{fossumEPL2006}. In the present study, the conductivity of dehydrated Na-FH clay particles that are used to prepare the ER suspensions is not known. Note that the conductivity of the suspending oil might also increase under application of an external electric field \cite{tangJRheology1995,foulcJElectrostat94}, and this may lead to a significant effect on the ER behavior.  

The CSR tests, on the contrary, provide a dynamic yield stress, $\tau_0^\text{d}$, as the extrapolation of the flow curves at low shear rates. It scales with respect to the applied field as a power law with an exponent $\alpha$ closer to $2$ than that obtained from the static and bifurcation yield stresses. This is consistent with previous  observation on data from laponite suspensions. Though the meaning of the dynamic yield stress in terms of yielding behavior of the suspensions is not clear, it is remarkable that, as for the laponite ER suspensions, we be able to scale the flow curves corresponding to a given particle fraction $\Phi$ so as to position them on the same master curve, following a scaling law in the form
\begin{equation}
\tau_d(\Phi,E) = E^\alpha f\left ( \frac{\dot{\gamma}}{E^2} \right ) \text{ ~.}
\end{equation}
In contrast to the CSR flow curves from laponite suspensions, though, the CSR flow curves from Na-FH suspension could not be scaled as a function of particle volume fractions (following Eq.\eqref{eq:scaling_law}). We attribute this to the large uncertainties connected to the real $\Phi$-values (sedimentation).

\subsubsection{ER effect of Na-fluorohectorite suspensions vs. ER effect of laponite suspensions}

Though (i) NaFH crystallites have a larger structural surface charge density ($1.2$e$^-/$unit cell) than laponite crystallites 
($0.4$e$^-/$unit cell), and (ii) the anisotropic shape of NaFH particles may cause enhanced dipolar interactions as compared to isotropic spherical particles of same size and same conductivity \cite{KanuJRheol98}, the measurements presented in the present study show that the ER effect results in a weaker yield stress fluid in the case of fluorohectorite particles than what was observed with laponite particles. We believe that the size and shape polydispersity of the clay particle population potentially makes the ER aggregates weaker, leading therefore to a reduction in shear stress- and viscosity- as these will depend on the packing of individual clay particles in chain or- column like-structures \cite{brooksColloidsSurf1986}.

\begin{table}
\centering
 \caption{\label{tab:exponent_comparison} Exponent $\alpha$ and $\beta$, as defined by Eq.~\eqref{eq:scaling_law}), obtained for ER laponite suspensions (results taken from ref.\cite{parmarLangmuir2008}) and NaFH ER suspensions (this study) from our three yield stress estimates: dynamic-, static-, and bifurcation- yield stress.}
\begin{tabular}{l || c | c || c | c}
\hline
\hline
& \multicolumn{2}{c||}{Laponite} & \multicolumn{2}{c}{Na-fluorohectorite} \\
\hline
& $\alpha$ & $\beta$ & $\alpha$ & $\beta$ \\
\hline
\hline
Dynamic y. s.    & 1.85 & 1.00 & 1.93 & --\\
Static y. s.      & 1.85 & 1.70 & 1.58 & 0.54\\
Bifurcation y. s. & 1.84 & 1.70 & \textit{0.5-1.6} & --\\
\hline
\hline
\end{tabular}
\end{table}
In terms of exponents of the scaling law, the $\alpha$ values are similar for both systems, for all estimates of the yield stress (see table~\ref{tab:exponent_comparison}). The $\beta$ values for fluorohectorite suspensions with $\Phi \leq 37.2$, on the contrary, are significantly lower than what was observed for laponite suspensions. In the latter ER fluids, the ``particles'' consist of aggregates of $1$nm-thick laponite platelets. From our prevous study \cite{parmarLangmuir2008}, we concluded that they are three-dimensional in nature. Here, prior to application of the electric field, the fluorohectorite particles are also aggregates of clay crystallites that are quasi-two-dimensional (being nano-stacks of average aspect ratio 10:1), but the crystallite population is very polydisperse, which probably leads to much larger shape and size variations in the clay aggregates in the oil. It is possible that at sufficiently large volume fraction, the aggregates become quasi-spherical in a way similar to those for laponite. This could explain the change of behavior observed for the largest particle fraction in Fig.~\ref{fig:figure6}.

\section{\label{sec:Conclusion and prospects} Conclusion and prospects}

We have studied the electrorheological behavior of ER fluids consisting of sodium fluorohectorite clay particles suspended in silicone oil. In the absence of electric field, these suspensions are Newtonian-like, and their viscosity is controlled by the particle volume fraction. Under application of a DC electric field larger than about 0.4 kV/mm, the flow curves are Bingham-like, exhibiting a well-defined yield stress that increases with increasing volume fraction of particles as well as with increasing applied electric field. The static and bifurcation yield stresses observed follow a power law $E^\alpha$, with $\alpha\sim1.5 - 1.6$ suggesting that interaction forces are governed by the conductivity mismatch between dipolar particles (fluorohectorite) and the suspending medium (silicone oil). 
The observed yield stresses are comparable to similar systems (see table 1 in ref.~\cite{parmarLangmuir2008}).

This works complements our previous study on these systems \cite{fossumEPL2006} with mechanical behaviors. The presently-obtained values for yield stresses suggest that such fluorohectorite-based ER fluids would not be able to compete with other systems in terms of the  practical uses of ER fluids. Rather, one application of these systems of (possibly functionalized) fluorohectorite particles could be in guided self-assemblies of nanoparticles for nano-templating and/or inclusion in composite materials.

In order to discriminate the effects of collective properties (geometry and cohesion of the packings) from that of individual particle properties (physical shape and chemical polarisibility) on the suspension's electrorheology, controlling the polydispersity of the suspensions and systematically studying samples based on several types of clay minerals is necessary, which also implies being able to measure aggregate size distributions in situ, in the oil. Such a systematic investigation is beyond the present study. Another prospect is the functionalization of the clay surfaces in order to ease single grain/nano-stack dispersion in oils. 

\section*{Acknowledgments}

This work was supported by NTNU, and by the Research Council of Norway (RCN) through the RCN NANOMAT Program) as well through a RCN SUP (Strategic University Program) project awarded to the COMPLEX Collaborative Research Team in Norway. Y. M. acknowledges the Egide organization for financial support in traveling between France and Norway, under the Aurora framework (grant nr. 18810WC). \additions{J. O. F. acknowledges travel support 
from the RCN under the same Aurora framework.}

\bibliographystyle{elsarticle-harv}
\bibliography{complete_new}







\end{document}